\documentclass[a4paper]{article} 
\renewcommand{\d}{{\mathrm d} } 
\newcommand{\nuc}[2]{{$^{#1}${#2}} } 
\usepackage{graphicx}
\begin{document}

\begin{center}
{\large\bf Nuclear vorticity and the low-energy nuclear response 
 - Towards the neutron drip line 
}\footnote{Work supported in part by 
Deutsche Forschungsgemeinschaft within the SFB 634 
and by the University of Athens under grant 70/4/3309.} 

\bigskip  
{P.~Papakonstantinou}$^{a,b}$,
{J.~Wambach}$^a$, 
{E.~Mavrommatis}$^b$, 
{V.Yu.~Ponomarev$^{a,}$\footnote{Permanent address: 
JINR, Dubna, Russia.} 
} 

\bigskip 
{\em 
$^a$Institut f\"ur Kernphysik, 
Technische Universit\"at Darmstadt,  
Schlossgartenstr.9, 
D-64289 
Darmstadt, Germany  
\\  
$^b$Physics Department, Nuclear and Particle 
Physics Section, University of Athens, 
GR-15771 Athens, Greece 
} 

\end{center} 



   
\begin{abstract} 
The transition density and current 
provide valuable insight 
into the nature of nuclear vibrations. 
   Nuclear vorticity is a quantity 
   related to the transverse transition 
   current. 
   In this work, we study the evolution of 
   the strength distribution, 
related to density fluctuations,  
   and 
   the vorticity strength distribution,  
   as the neutron drip line is approached. 
   Our results on the 
   isoscalar, 
   natural-parity multipole response 
   of Ni isotopes, obtained by using a 
   self-consistent Skyrme-Hartree-Fock + Continuum RPA model, 
   indicate that, 
close to the drip line, 
   the low-energy response is 
dominated by $L>1$ vortical transitions. 


\end{abstract} 

%

%

\section{Introduction} 

One of the aims of the ongoing theoretical and experimental 
studies devoted to nuclei far from stability 
is to illuminate 
the effects of the excess protons or neutrons on the 
nuclear structure and response. 
On the neutron-rich side of the nuclear chart, 
in particular,  
a wealth of exotic phenomena 
is anticipated, 
due to the formation of neutron 
skins and halos, the large 
difference between the Fermi energies 
of protons and neutrons 
and a possible change of shell structure. 

Electron scattering experiments with 
radioactive beams are currently being discussed, opening 
the possibility of detecting, 
besides the transition charge distributions, 
also transition current distributions in a 
broad region of the nuclear chart. 
The transition current density (TC), studied in conjunction with 
the transition density (TD),  
provides valuable insight 
into the nature of nuclear vibrations, 
in particular their ``zero-sound" nature. 
Theoretically, the TD and TC 
associated with a particular 
type of electric excitation, 
approach ideal hydrodynamical behaviour 
in the case of very collective states such as 
giant resonances (GRs). It is 
predicted~\cite{HSZ1999}  that 
the properties of GRs 
in drip-line nuclei are 
close to the ones of $\beta -$stable nuclei 
in this sense. 
At lower energies, though, 
the behaviour of the TD and TC 
may deviate significantly from this picture, 
as several examples demonstrate. 
Microscopic studies of convection currents in 
stable nuclei 
predict that a variety of flow patterns is 
possible 
\cite{SDS1983,WMK1983,RaW1987};  
for instance, the isoscalar quadrupole 
GR of \nuc{208}{Pb} 
is nearly irrotational,   
whereas the corresponding 
low-lying state exhibits 
considerable vorticity. 
In (stable and unstable) neutron-rich nuclei, 
``pygmy" \cite{Har2000,Rye2002} 
or ``soft" \cite{Le01,Tr02,Nak2000} 
dipole modes are observed, interpreted as 
oscillations of the excess neutrons 
against the isospin-saturated core;  
also, toroidal dipole modes are predicted to exist at a somewhat 
higher energy
\cite{Rye2002,VPR2002}. 
A special  feature of the 
response of 
very neutron-rich nuclei is the so-called 
threshold strength, 
i.e. 
the considerable amount of strength predicted to occur  
just above the neutron threshold; it 
is 
of 
isoscalar and non-collective nature and characterized by 
spatially extended 
TDs and TCs, 
owing to the 
loosely bound neutrons 
\cite{HSZ1999,HSZ1996,HaS1996b,HSZ1997b,CLN1997b}. 

The structure of the particle continuum 
and the nature of low-lying strength 
in unstable nuclei 
are of particular interest,  
as they provide information on 
 decay properties, 
polarizabilities, shell structure etc.  
Close to the drip line, 
collective ``low-lying" states such as $2^+$ 
may be located in the continuum. 
The energy of $2^+$ states as well as 
other $0\hbar\omega$ modes, 
may be very low even for nuclei described 
traditionally as closed-subshell 
configurations, 
if 
the shell structure 
tends to ``melt" at the extremes of isospin \cite{HSZ1996}. 
Such  transitions 
can hardly be expected to 
be irrotational. 

A standard procedure in studies of nuclear convection 
currents is to define a velocity field, as 
dictated by hydrodynamical considerations. 
In Ref.~\cite{RaW1987}, it was shown that 
the degree to which a transition 
is irrotational,  
can be quantified in terms of 
the nuclear vorticity density. 
The latter is directly related to the 
transverse component of the TC 
and therefore, together with the TD, 
it completely specifies a transition 
and unambiguously reflects the zero-sound character of 
nuclear vibrations. 

In this work, we examine 
the response of the isotopes 
\nuc{56,78,110}{Ni},  
ranging from $N=Z$ and 
close-to-stable ($N=28$) 
to extremely neutron-rich ($N=82$).  
We focus on the 
flow properties of the various transitions. 
The isoscalar strength distributions 
$S_{\rho}(E)$, 
related to density fluctuations,  and 
respective vorticity strength distributions $S_{\omega}(E)$ 
of multipolarity up to 
$L=4$ 
are calculated, 
using a self-consistent Skyrme-Hartree-Fock 
plus Continuum-RPA method. 
We consider the vorticity related to the convection current. 
The evolution of the distributions 
as the neutron number increases is discussed. 

\section{Definitions and method of calculation} 
Let us consider the response of 
spherical, closed-(sub)shell nuclei 
to an isoscalar (IS) external field 
$
V_{LM}(\vec{r}) = r^LY_{LM}(\theta,\phi) 
$ 
($L>0$).  
The IS TD 
$\delta\rho (\vec{r})$ 
and TC  
$\vec{j}(\vec{r})$ 
characterizing 
the excited natural-parity state $|LM\rangle$ 
of energy $E$, 
are determined 
by their radial components, 
$\delta\rho_L(r)$ and $j_{LL\pm 1}(r)$ 
respectively \cite{SDS1983,RaW1987}. 
From 
$\vec{j}(\vec{r})$ and 
the ground-state density $\rho (r)$, 
a velocity field 
can be defined, 
$ 
\vec{u}(\vec{r}) = \vec{j}(\vec{r})/\rho(r) 
$. 
Under the assumption that a single transition exhausts 
the available strength of an external potential 
$V(\vec{r})$, it 
can be shown that 
$\vec{u}(\vec{r}) \propto \vec{\nabla}V(\vec{r}) 
$ \cite{Su84}, 
which implies irrotational flow, 
$  
\vec{\nabla}\times\vec{u} =0 
$. 
In classical hydrodynamics, the flow is always irrotational in 
ideal, non-viscous fluids. 

The IS TD and TC 
are related through the 
continuity equation (CE) \cite{SDS1983}, 
which we write here for the radial 
components: 
\begin{equation} 
\frac{E}{\hbar c} 
\delta\rho_L(r) 
 =  
\sqrt{\frac{L}{2L+1}}(\frac{\d}{\d r}-\frac{L-1}{r})j_{LL-1}(r) 
-\sqrt{\frac{L+1}{2L+1}}(\frac{\d}{\d r}+\frac{L+2}{r})j_{LL+1}(r) 
\label{Eceq} 
 . 
\end{equation} 
The transverse part of the current, $\vec{j}^{\mathrm{tr}}$, 
with radial components $j^{\mathrm{tr}}_{LL\pm 1}(r)$, 
does not contribute to the CE since, by definition, 
$ 
\vec{\nabla}\cdot\vec{j}^{\mathrm{tr}} = 0 
$. 
This equation does not define $\vec{j}^{\mathrm{tr}}$ 
uniquely; following Ref.~\cite{HLP1982}, 
one may set $j^{\mathrm{tr}}_{LL+1}(r)=j_{LL+1}(r)$. 
Then $j^{\mathrm{tr}}_{LL-1}(r)$ is 
determined from 
$\vec{\nabla}\cdot\vec{j}^{\mathrm{tr}} = 0$.  
The remaining part of the current 
is related to $\delta\rho_L$ through the CE.  
Therefore, measurements of $\delta\rho_L$ 
and $j_{LL+1}(r)$, 
as demonstrated for example in the analyses of 
inelastic-electron-scattering data in Ref.~\cite{HLP1982}, 
determine the transition completely. 

The {vorticity} density $\vec{\omega}(\vec{r})$ 
associated with a transition 
from the 
spherically symmetric ground state 
to the 
excited state $|LM\rangle$, as  
defined in Ref.~\cite{RaW1987}, %
equals the curl of the transverse current 
introduced above \cite{Cad1999}, 
\begin{equation} 
\vec{\omega}(\vec{r}) = 
\vec{\nabla}\times
\vec{j}^{\mathrm{tr}}(\vec{r})  
\equiv  
(2L+1)^{-1/2} 
 \omega_{LL}(r)\vec{Y}_{LL}^{M}(\hat{r}) 
 . 
\end{equation} 
The radial part $\omega_{LL}(r)$ is given by 
\begin{equation} 
    \omega_{LL}(r) 
 = \sqrt{\frac{2L+1}{L}} (\frac{\d}{\d r} + \frac{L+2}{r}) j_{L L+1}(r) 
\label{Evor} 
 . 
\end{equation} 
The $r^L-$ moment of $\omega_{LL}$ 
is identically zero; 
its $r^{L+2}-$moment may be used as a measure of its 
strength \cite{RaW1987}. 
For irrotational and incompressible flow, 
e.g. in the Tassie model, 
$\omega_{LL}$ vanishes. 
 
The strength distribution 
$S_{\rho}(E)=\sum_f|\langle f |{V}_{LM}|0\rangle |^2 
\delta (E-E_f)$, 
where $|f\rangle$ are the final states excited by the 
field ${V}_{LM}$ and $E_f$ their excitation energies, 
is related to the 
$r^L-$moment of $\delta\rho_L(r)$. 
Since we are 
dealing with continuous distributions, 
we write the strength in a small energy 
interval of width $\Delta E$ at energy $E$ as 
$ 
S_{\rho}(E) 
= [\int_0^{\infty} \d r \delta\rho_L(r) r^{L+2}]^2  
 / \Delta E 
$. 
We define the {\em vorticity} strength distribution 
$S_{\omega}(E)$ in a similar way, 
namely 
$ 
S_{\omega}(E) = 
[\int_0^{\infty} \d r \omega_{LL} (r) r^{L+4}]^2  
 / \Delta E 
$. 
Comparison of 
$ 
S_{\rho}(E) 
$  
and 
$ 
S_{\omega}(E) 
$   
allows a discrimination 
between collective states of longitudinal 
zero-sound character  
and excitation modes which are highly vortical 
and possibly of transverse zero-sound character. 
The latter case applies, for example, to the toroidal 
dipole mode 
\cite{Rye2002,VPR2002}. 
Note that, in order to quantify how 
``collective" (in terms of $S_{\omega}$)  
a seemingly strong 
transition is, and to what extent it is a  
transverse-zero-sound candidate, one would need to 
establish 
appropriate 
sum rules for $S_{\omega}$. 
%
%
%

The quantities introduced above 
are calculated 
using a 
Skyrme - Hartree-Fock (HF) plus Continuum - RPA model. 
For the HF ground - state, 
the numerical code of P.-G.~Reinhard \cite{ReXX} 
is used. 
The calculation of 
the response function 
is formulated 
in coordinate space, 
as described in 
 \cite{BeXX,BeT1975,vGi1983,Ryc1988}. 
First, for given energy $E$, multipolarity $L$ and 
isospin charachter $\tau_z$, 
the radial part 
$G_{\mu_1\mu_2}^0(r,r';E)  
$ 
of the unperturbed $ph$ Green function $G^0(E)$ 
is calculated. 
The indices  
$\mu_i = 1, 2, \ldots $ enumerate the operators 
whose propagation is described by the Green function, 
namely 
\[ Y_{LM}, \,\,   
 [Y_L\otimes (\nabla^2+{\nabla '}^2) ]_{LM}, \,\,    
 [Y_{L \pm 1}\otimes (\vec{\nabla} \pm \vec{\nabla '})]_{LM}  
. 
\]   
Spin-dependent terms have been omitted in the present calculation. 
The continuum is fully taken into account, 
as described in \cite{BeXX,vGi1983}. 
A small but finite Im$E$ ensures that bound transitions 
acquire a finite width \cite{BeXX}. 
The RPA $ph$ Green function is obtained by 
solving 
the equation 
\begin{equation} 
G^{\mathrm{RPA}}(E) = [ 1+G^0(E)V_{\mathrm{res}}]^{-1} 
G^0(E) 
\end{equation} 
as a matrix equation in 
coordinate space (represented by a radial mesh),  
isospin character 
and 
operators $\mu_i$. 
The $ph$ {residual interaction} 
$V_{\mathrm{res}}$ is 
derived self-consistently from the Skyrme-HF energy 
functional \cite{vGi1983,Ryc1988,Tsa1978}.   
The radial functions 
$\delta\rho_L$, $j_{LL-1}$ and $j_{LL+1}$, 
at given energy $E$,  
can be calculated from the 
RPA Green function;  
then, $\omega_{LL}$ is obtained 
using Eq.~(\ref{Evor}) 
\cite{Pa04,PWPhr}.  
 
%
%
\section{Results and discussion} 

We have performed calculations for the IS natural-parity 
response of Ni isotopes 
for $L\leq 4$. 
Since pairing is not included in our calculation scheme, 
only the doubly closed isotopes 
\nuc{56}{Ni} ($N=28$),  
\nuc{78}{Ni} ($N=50$) and   
\nuc{110}{Ni} ($N=82$) have been studied. 
The nucleus \nuc{56}{Ni}  
lies close to the valley of stability, 
while  
\nuc{110}{Ni} serves as a numerical example of 
a nucleus with extreme neutron excess 
lying close to the drip line. 
In all cases we examined,  
we have checked that the CE, Eq.~(\ref{Eceq}), 
is fulfilled to a satisfactory degree. 
We present results 
for the IS 
$2^+$, $3^-$  and $4^+$ 
strength distribution $S_{\rho}(E)$ 
and the 
corresponding 
vorticity strength distribution $S_{\omega}(E)$.  
The distributions are normalized to 1, i.e. 
$S_{\rho ,\omega}(E)/S_{\rho ,\omega ,\,\,\mathrm{ tot}}$ 
is shown, where $S_{\rho ,\omega ,\,\,\mathrm{ tot}}$ 
is the integrated strength in the region 0-50~MeV. 
For the results shown we have used 
the Skyrme-force parametrization {SkM*} \cite{Bar1982}, 
tailored to describe GRs of 
stable nuclei and 
employed in previous studies of 
the response of exotic nuclei as well  
(eg. in \cite{HSZ1999,HSZ1996,HaS1996b,HSZ1997b}). 

As we observe in Fig.~1, 
most of the IS quadrupole strength is concentrated in two peaks, 
a low-energy collective state and the GR. 
The width of the GR increases and its energy 
decreases as we approach the neutron drip line. 
The GR 
carries a large fraction of $S_{\rho ,\,\, \mathrm{ tot}}$ 
and 
little 
vorticity. It has been verified that, 
in the three Ni isotopes studied, 
its 
velocity field resembles that of 
ideal hydrodynamical flow.  
On the contrary, the  
lowest $2^+$ state is strongly vortical. 
The octupole response of \nuc{56}{Ni} (\nuc{78}{Ni}), 
Fig.~2, 
is dominated by one (two) 
collective state(s) carrying 
almost no vorticity strength; 
a large part of the $S_{\omega}$ 
lies in the $3\hbar\omega$ region. 
The situation is quite different in the case of 
\nuc{110}{Ni}, 
where 
highly vortical 
threshold 
strength appears at low energy. 
In order to illustrate the 
difference, in Fig.~3 we plot the velocity field corresponding 
to the 9.3~MeV state in \nuc{56}{Ni} and 
the 3.4~MeV transition in \nuc{110}{Ni}, 
in an arbitrary scale. 
For \nuc{56,78}{Ni}, it is interesting to notice a 
very weak octupole state with considerable vorticity strength, 
at around 15~MeV. 
The hexadecapole strength distribution 
of \nuc{56,78}{Ni}, Fig.~4, 
appears fragmented. Vorticity strength is carried 
by all peaks. In 
\nuc{110}{Ni} 
a striking amount of vorticity 
develops at very low energy. 

The neutron-rich nucleus 
\nuc{78}{Ni} 
is possibly doubly magic \cite{Da00}. 
We notice that its low-energy response 
does not differ dramatically from 
the response of \nuc{56}{Ni}. 
The transition 
strength of \nuc{110}{Ni} is located in the continuum 
in all examined 
cases. 
We have performed calculations also for 
the $0^+$ and $1^-$ response of the 
three isotopes. In \nuc{110}{Ni}, 
the threshold energy for these multipoles is 
higher than for the $L=2,3,4$ multipoles. 
The threshold strength 
of \nuc{110}{Ni} for $L>0$ carries 
increased vortical strength. 
This result should be anticipated  given the  
single -
particle 
character 
of the threshold strength.   
The present results indicate that, 
close to the neutron drip line, 
the low-energy response is dominated by 
$L>1$ transitions 
of transverse character. 
$1^-$ as well as $3^-$ excitations 
may gain importance with respect to $2^+$ and $4^+$ 
in the special 
case of nuclei described as $n\ell -$closed 
configurations. 

We have also 
used the Skyrme force  
{MSk7} \cite{GTP2001}, whose parameters were 
determined 
by fitting the values of nuclear masses, calculated using  
the HF+BCS method, to the measured  
ones, for 1888 nuclei 
with various values of $|N-Z|/A$. 
The two forces have similar 
nuclear-matter properties, except for the 
effective mass $m^{\ast}$. 
The SkM* force ($m^{\ast}/m = 0.786$) 
predicts that the last occupied neutron 
state of \nuc{110}{Ni} is bound. 
The nucleus appears soft 
against excitations of vortical nature. 
In the case of the MSk7 force 
($m^{\ast}/m=1.05$),  
the last neutron state of \nuc{110}{Ni} 
appears as a resonant state at positive energy 
and therefore the nucleus lies beyond the neutron 
drip line.  
In general, the results obtained with the 
two Skyrme parametrizations agree qualitatively. 
The strength distributions are 
systematically shifted towards lower energies 
when MSk7 is used. 

We should stress that the nucleus \nuc{110}{Ni} was studied 
as an academic example of a closed-shell Ni isotope 
close to the neutron drip line, as has been done before 
in Ref.~\cite{HSZ1996}. 
This particular choice was dictated by the limitations of our model, 
which does not account for pairing correlations. 
According to SHF results, the $N=82$ closure may still 
be valid in the Ni region \cite{HSZ1996}, although 
not conclusively. 
Ideally, one should perform a (preferably self-consistent) 
Quasiparticle-RPA (QRPA) calculation including the full continuum, 
as proposed in Ref.~\cite{Mat2001}. 
The correct treatment of the continuum when pairing is present, 
in the ground state as well as in the excited states,  
is particularly important 
in weakly bound neutron systems, where 
the pairing field couples the neutrons in the 
bound states with those in the low-energy 
continuum \cite{BBD2002,HaM2003,KSG2002}.  
In general, pairing correlations 
are not expected to 
modify dramatically the electric transitions of even-even 
nuclei (the same may not hold for magnetic transitions, 
however) \cite{Ro70}. They can introduce new transitions and 
change the energy and strength of the ones 
that we have already examined \cite{Mat2001,BM75}. 
Therefore, we expect that our 
conclusions would still hold qualitatively, after pairing is 
taken into account. Of course, they are restricted 
to spherical nuclei. 

In conclusion, 
   our results for the 
   IS, 
   natural-parity response 
   of Ni isotopes, obtained using a 
   self-consistent Skyrme-Hartree-Fock + Continuum RPA model, 
   indicate that, 
close to the neutron drip line, 
instabilities develop in 
the form of 
$L>1$ transitions of transverse 
character. 
We have used the vorticity strength distribution 
to quantify the irrotational character of 
the various transitions.  
Future  work should account for the 
spin current, besides the convection current. 
It would be useful to 
examine the magnetic response as well, 
for a more complete picture. 
Ideally, a chain of even$-N$ isotopes should be 
studied using a QRPA method including the continuum.  
The proton drip line is also interesting to 
explore; modifications of the 
strength distributions, in the direction 
of sharper low-lying transitions, 
are expected due to the Coulomb barrier. 

\subsection*{Acknowledgements} 
P.P. is grateful to Prof.~K.~Heyde for the kind 
hospitality at the Department of Subatomic and Radiation Physics, 
University of Gent, Belgium, 
and wishes to thank, 
in particular, Drs. N.~Jachowicz, J.~Ryckebusch 
and D.~Van~Neck 
for valuable help with the Continuum-RPA method. 

%
%

\begin{thebibliography}{10}

\bibitem{HSZ1999}
I.~Hamamoto, H.~Sagawa and X.Z. Zhang,
\newblock { Nucl. Phys. { A 648}} (1999) 
\newblock 203.

\bibitem{SDS1983}
F.E. Serr, T.S. Dimitrescu, T.~Suzuki and C.H.Dasso,
\newblock { Nucl. Phys. { A 404}} (1983) 
\newblock 359.

\bibitem{WMK1983}
E.~W{\"u}st, U.~Mosel, J.~Kunz and A.~Schuh,
\newblock { Nucl. Phys. { A 406}} (1983) 
\newblock 285.

\bibitem{RaW1987}
D.G. Ravenhall and J.~Wambach,
\newblock { Nucl. Phys. { A 475}} (1987) 
\newblock 468.

\bibitem{Har2000}
T.~Hartmann, J.~Enders, P.~Mohr, K.~Vogt, S.~Volz and A.~Zilges,
\newblock { Phys. Rev. Lett. { 85}} (2001) 
\newblock 274.

\bibitem{Rye2002}
N.~Ryezayeva et~al, 
\newblock { Phys. Rev. Lett. { 89}} (2002) 
\newblock 272502.

\bibitem{Le01}
A.~Leistenschneider et~al,
\newblock { Phys. Rev. Lett. { 86}} (2001) 
\newblock 5442.

\bibitem{Tr02}
E.~Tryggestad et~al,
\newblock { Phys. Lett. { B 541}} (2002) 
\newblock 52.

\bibitem{Nak2000}
S.~Nakayama et~al,
\newblock { Phys. Rev. Lett. { 85}} (2000) 
\newblock 262.

\bibitem{VPR2002}
D.~Vretenar, N.~Paar, P.~Ring and T.~Niksic,
\newblock { Phys. Rev. { C 65}} (2002) 
\newblock 021301(R).  

\bibitem{HSZ1996}
I.~Hamamoto, H.~Sagawa and X.Z. Zhang,
\newblock { Phys. Rev. { C 53}} (1996) 
\newblock 765.

\bibitem{HaS1996b}
I.~Hamamoto and H.~Sagawa,
\newblock { Phys. Rev. { C 54}} (1996) 
\newblock 2369.

\bibitem{HSZ1997b}
I.~Hamamoto, H.~Sagawa and X.Z. Zhang,
\newblock { Phys. Rev. { C 56}} (1997) 
\newblock 3121.

\bibitem{CLN1997b}
F.~Catara, E.G. Lanza, M.A. Nagarajan and A.~Vitturi,
\newblock { Nucl. Phys. { A 624}} (1997) 
\newblock 449.

\bibitem{Su84}
T.~Suzuki,
\newblock { Ann. Phys. Fr. { 9}} (1984) 
\newblock 535.

\bibitem{HLP1982}
J.~Heisenberg, J.~Lichtenstadt, C.N. Papanicolas and J.S. McCarthy,
\newblock { Phys. Rev. { C 25}} (1982) 
\newblock 2292.

\bibitem{Cad1999}
E.C. Caparelli and E.J.V. de~Passos,
\newblock { J. Phys. { G 25}} (1999) 
\newblock 537.

\bibitem{ReXX}
P.-G. Reinhard,
\newblock {Skyrme-Hartree-Fock Model},
\newblock { {in: Computational Nuclear Physics I - Nuclear Structure}}, eds.
  K. Langanke, J.E. Maruhn and S.E. Koonin (Springer, New York 1991) 
\newblock p.28.

\bibitem{BeXX}
G.~Bertsch,
\newblock The Random Phase Approximation for Collective Excitations,
\newblock { {ibid, 
}} 
\newblock p.75.

\bibitem{BeT1975}
G.F. Bertsch and S.F. Tsai,
\newblock { Phys. Rep. { 18 C}} (1975) 
\newblock 125.

\bibitem{vGi1983}
N.~Van Giai, 
\newblock {Nuclear excitations in the nuclear response theory}, 
\newblock {in: Nuclear Collective Dynamics}, 
  eds. D.~Bucuresku, V.~Ceausescu, N.V.~Zamfir 
  (World Scientific, 1983) p.356. 

\bibitem{Ryc1988}
J.~Ryckebusch,
\newblock Ph.D. thesis, University of Gent, 1988.

\bibitem{Tsa1978}
S.F. Tsai,
\newblock { Phys. Rev. { C 17}} (1978) 
\newblock 1862.

\bibitem{Pa04}
P.~Papakonstantinou,
\newblock Ph.D. thesis, University of Athens, 2004.

\bibitem{PWPhr}
P.~Papakonstantinou, J.~Wambach, E.~Mavrommatis and V.~Yu. Ponomarev,
\newblock {in: Proc. Int. Workshop XXXII on Gross Properties of Nuclei and
  Nuclear Excitations}, Hirschegg, Kleinwalsertal, Austria, 
  January 11-17, 2004;
\newblock p.~159. 

\bibitem{Bar1982}
J.~Bartel, P.~Quentin, M.~Brack, C.~Guet and {H.-B.} Hakansson,
\newblock { Nucl. Phys. { A 386}} (1982) 
\newblock 79.

\bibitem{Da00}
J.M. Daugas, R.~Grzywacz, M.~Lewitowicz, L.~Achouri et~al,
\newblock { Phys. Lett. { B 476}} (2000) 
\newblock 213.

\bibitem{GTP2001}
S.~Goriely, F.~Tondeur and J.M. Pearson,
\newblock { At. Dat. Nucl. Dat. Tables { 77}} (2001) 
\newblock 311.

\bibitem{Mat2001} 
M.~Matsuo, Nucl. Phys. {A 696} (2001) 371. 

\bibitem{BBD2002} 
K.~Bennaceur, J.F.~Berger and B.~Ducomet, 
\newblock {Nucl. Phys. A 708} (2002) 
\newblock 205. 

\bibitem{HaM2003} 
I.~Hamamoto and B.R.~Mottelson, 
\newblock Phys. Rev. C 68 (2003) 
\newblock 034312. 

\bibitem{KSG2002} 
E.~Khan, N.~Sandulescu, M.~Grasso and N.~Van Giai, 
\newblock Phys. Rev. C 66 (2002) 
\newblock 024309. 

\bibitem{Ro70} 
D.J.~Rowe, ``Nuclear Collective Motion", Methuen 
and Co. Ltd. 1970. 

\bibitem{BM75} 
A.~Bohr and B.R.~Mottelson, ``Nuclear Structure", Vol II, 
Benjamin, 1975. 


\end{thebibliography}

\clearpage 

%
\begin{figure} 
\begin{center}
   \includegraphics[width=8.0cm]{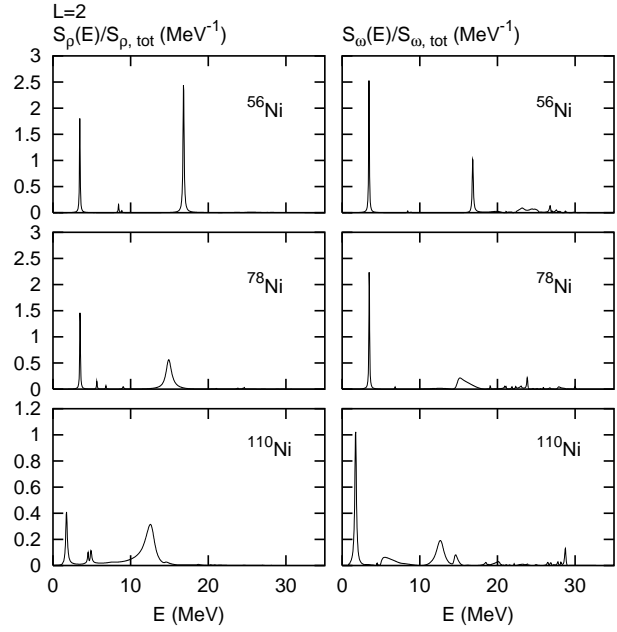} 
\end{center}
     \caption{%
Quadrupole strength distribution (left) 
and vorticity-strength distribution (right), 
normalized to 1, for the isotopes 
\nuc{56,78,110}{Ni}. The Skyrme parametrization 
SkM* has been used.%
} 
\end{figure} 
\begin{figure} 
\begin{center}
   \includegraphics[width=8.0cm]{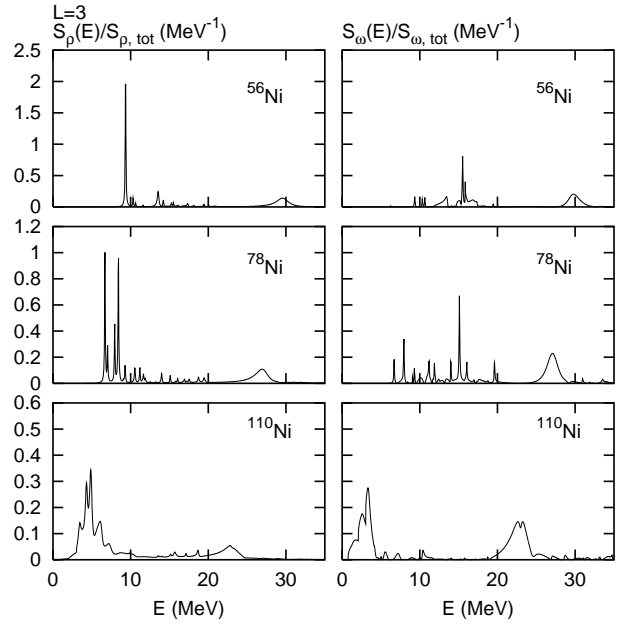} 
\end{center}
     \caption{%
As in Fig.~1, 
octupole strength distributions. 
} 
\end{figure} 
\begin{figure} 
\begin{center}
   \includegraphics[angle = 90, width=8.0cm]{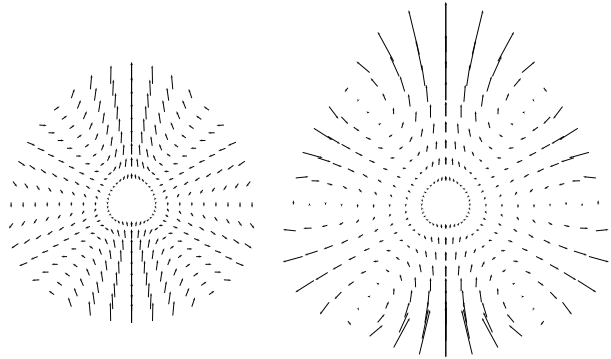}\\
\end{center}
     \caption{%
Left (Right): 
Velocity field for the octupole excitation 
of 
\nuc{56}{Ni} at $E=9.3$~MeV 
(\nuc{110}{Ni} at $E=3.4$~MeV) 
plotted up to 
a radial distance of 6~fm (7.6~fm). 
The scale of the velocity 
amplitude is arbitrary. The $z-$axis is along the page.} 
\end{figure} 
\begin{figure} 
\begin{center}
   \includegraphics[width=8.0cm]{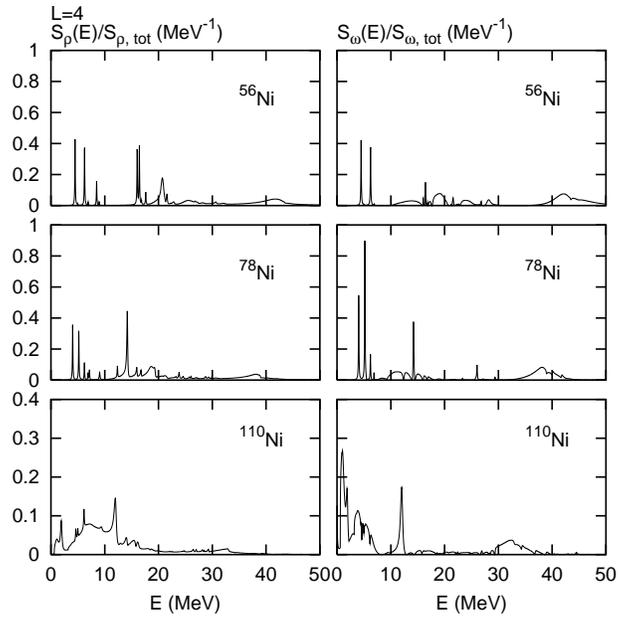} 
\end{center}
     \caption{%
As in Fig.~1, 
hexadecapole strength distributions. 
} 
\end{figure} 

\end{document}